\def\CHCF{$\beta^{\prime\prime}$-(BEDT\--TTF)$_2$\-SF$_5$\-CH$_2$\-CF$_2$\-SO$_3$}
\def\d8CHCF{$\beta^{\prime\prime}$-(d$_8$-BEDT\--TTF)$_2$\-SF$_5$\-CH$_2$CF$_2$\-SO$_3$}
\def\CHF{$\beta^{\prime\prime}$-(BEDT\--TTF)$_2$\-SF$_5$\-CHF\-SO$_3$}
\documentclass[epj]{svjour}
\usepackage{graphics}
\begin{document}

\title{DC and high-frequency conductivity of the organic metals
$\beta^{\prime\prime}$-(BEDT-TTF)$_2$SF$_5R$\,SO$_3$ ($R$ = CH$_2$CF$_2$ and CHF)
}
\author{M. Glied\inst{1}
\and S. Yasin\inst{1} \and S. Kaiser\inst{1} \and N.
Drichko\inst{1} \and M. Dressel\inst{1}
\thanks{email: dressel@pi1.physik.uni-stuttgart.de}
\and J. Wosnitza\inst{2}
\and J.A.  Schlueter\inst{3} \and G.L. Gard\inst{4} }

\institute{1. Physikalisches Institut, Universit\"at Stuttgart,
Pfaffenwaldring 57, D-70550 Stuttgart, Germany \and
Dresden High-Magnetic Field Laboratory, Forschungszentrum Dresden-Rossendorf, D-01314 Dresden, Germany \and
Material Science Division, Argonne National Laboratory, Argonne,
Illinois 60439-4831, U.S.A.
\and
Department of Chemistry, Portland
State University, Portland, Oregon 97207-0751, U.S.A.}

\date{Received: \today}

\abstract{The temperature dependences of the electric-transport
properties of the two-dimensional organic conductors  \CHCF, \d8CHCF,
and \CHF\ are measured by dc methods in and perpendicular to the
highly-conducting plane. Microwave measurements are performed at
24 and 33.5~GHz to probe the high-frequency behavior
from room temperature down to 2~K.
Superconductivity is observed in \CHCF\ and its deuterated
analogue. Although all the compounds remain metallic down to
low-temperatures, they are close to a charge-order transition.
This leads to deviations from a simple Drude behavior of the
optical conductivity which
become obvious already in the microwave range. In \CHCF, for
instance, charge  fluctuations cause an increase in microwave
resistivity for $T<20$~K which is not detected in dc measurements.
\CHF\ exhibits a simple metallic behavior at all
frequencies. In the dc transport, however, we observe
indications of localization  in the perpendicular direction.
}
\titlerunning{High-frequency conductivity of the organic metals}

\maketitle

\section{Introduction}
Superconductivity has been discovered in organic crystals almost
three decades ago; nevertheless they are still considered as novel
materials with exotic properties because many questions remain
unanswered in spite of the enormous progress achieved over these
years \cite{Ishiguro98}. It became clear that the normal-state
properties are distinct from those of conventional metals and
severely influence the superconducting ground state. Several issues of
crucial importance for the understanding of low-dimensional
electron systems have been identified that are currently
investigated in few distinct families of organic conductors,
which serve as prototypes \cite{ChemicalReviews}.

The $\kappa$-phase BEDT-TTF salts (where BEDT-TTF or ET stands for
bis-(ethyl\-ene\-di\-thio)\-te\-tra\-thia\-ful\-va\-lene), for
instance, are two-dimensional metals with a half-filled conduction
band and a superconducting transition temperature $T_c$ of almost
13~K. The system is close to a metal-insulator transition driven
by electronic correlations $U$. At room temperature the transport
reflects more a semiconducting behavior, often described as a bad
metal, and only below 100~K coherent transport starts to emerge
with a narrow Drude-like contribution developing as the
temperature is reduced further
\cite{Kornelsen90,Faltermeier07,Dumm09}. By now there is a good
agreement between experiment and theory \cite{Merino08} that the
$\kappa$-salts represent a prime example of a band-width controlled Mott
metal-insulator transition in two dimensions, and the correlated carriers in
the narrow zero-frequency contribution of the optical conductivity
exhibit Fermi-liquid behavior.
\begin{figure}[b]
\centering\resizebox{0.4\textwidth}{!}{\includegraphics*{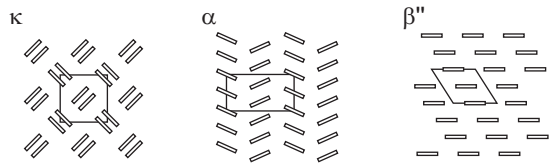}}
\caption{
\label{fig:pattern}
The BEDT-TTF molecules are arranged in different patterns for
the various phases.}
\end{figure}

The $\alpha$-phase BEDT-TTF compounds are quarter-filled systems,
which are subject to charge order depending on the ratio of
nearest-neighbor interaction $V$ to bandwidth $W$
\cite{Drichko07}: while $\alpha$-(BEDT-TTF)$_2$I$_3$ becomes
completely insulating below a sharp metal-insulator transition at
135~K \cite{Bender84,Dressel92a},
$\alpha$-(BEDT-TTF)$_2$\-(NH$_4$)\-Hg(SCN)$_4$ is an ambient
pressure superconductor with $T_c\approx 1$~K \cite{Wang90}. If in
the anions (NH)$_4$ is substituted by Tl, Rb, or K,
superconductivity is suppressed but metallic properties are
conserved all the way down to lowest temperatures. Accordingly,
the optical conductivity exhibits a narrow Drude-like component
within a charge-order pseudogap
\cite{Dressel92b,Dressel03,Drichko06}. It was proposed that in
these systems superconductivity is mediated by charge-order
fluctuations \cite{Merino01}, and recent experiments
\cite{Drichko06,Merino06} provide strong evidence pointing in this
direction.

In order to clarify how general this proposal is, other
quarter-filled compounds have to be investigated that also show
charge order, like the $\theta$ or $\beta^{\prime\prime}$ phases
of the BEDT-TTF salts.
In
the case of the $\beta^{\prime\prime}$ phase, there is only one
kind of stack with two BEDT-TTF molecules per unit cell \cite{Mori98a} as depicted in Fig.~\ref{fig:pattern}. The
planar BEDT-TTF molecules are loosely stacked along the $a$ axis
with strong orbital overlaps between adjacent stacks ($b$
direction). The highly conducting $ab$ sheets are separated by the
anions with a much lower conductivity in $c$ direction.

Recently optical investigations on
$\beta^{\prime\prime}$-(BEDT-TTF)\-(TCNQ) have been published
\cite{Uruichi06} which indicate a charge-order transition around
170~K. The interplay of charge order and superconductivity was
studied by magnetotransport experiments \cite{Bangura05} in a
series of compounds based on
$\beta^{\prime\prime}$-(BEDT-TTF)$_4$[(H$_3$O)$M$(C$_2$O$_4$)$_3$].
Here we will concentrate on the family
$\beta^{\prime\prime}$-(BEDT-TTF)$_2$\-SF$_5\-R\-$\,SO$_3$.

\CHCF\ is the first fully organic superconductor of the BEDT-TTF
family \cite{Geiser96}. If the anions are slightly modified by
replacing $R$ = CH$_2$CF$_2$ with CHF in
$\beta^{\prime\prime}$-(BEDT-TTF)$_2$SF$_5R$\,SO$_3$, the material
remains metallic down to low temperatures but does not
superconduct. Changing the anions to SF$_5$CHFCF$_2$SO$_3$ causes
a metal to insulator transition near 180~K, while crystals with
$R$ = CF$_2$ and CH$_2$ are insulators in the entire temperature
range \cite{Ward00,Jones00}. The question addressed here is the
nature of the charge transport, the effects of electronic
correlations, and the influence of charge fluctuations on
the electronic properties in the metallic
$\beta^{\prime\prime}$-(BEDT-TTF)$_2$SF$_5R$\,SO$_3$ salts. To
this end, we have performed dc and microwave measurements of
the in-plane resistivity as a function of temperature on the
organic superconductors \CHCF\ and deuterated analog \d8CHCF, as
well as the metallic compound \CHF.

\section{Experimental Details}
\label{sec:experimental} Single crystals of \CHCF, \d8CHCF, and
\CHF\ have been grown via electrochemical techniques in an H cell
at Argonne National Laboratory as described in
Ref.~\cite{Geiser96}. The crystals have typical dimensions of
$0.5\times 1.5 \times 0.25$~mm$^3$ and form plates with a large
face containing the conducting ($ab$) plane. We have performed
temperature-dependent measurements of the dc resistivity of these
crystals within the plane and perpendicular to it ($c$ direction)
using the standard four-point method with a typical current of
50~$\mu$A. The contacts were made by evaporating 50~\AA\ thick
gold pads on the crystal, then 15~$\mu$m gold wires were pasted on
each pad with a small amount of carbon paint; in some cases the
leads were directly put onto the crystal. The samples were slowly
(0.2 - 0.5~K/min) cooled down to avoid cracks and ensure thermal
equilibrium. Temperature-dependent measurements were conducted in
a He exchange gas cryostat down to 2~K; in general data were
acquired upon cooling and warming.

It is a well-known problem that the dc transport measurements of
highly-anisotropic samples are hampered by contributions from
other directions. This is particularly important when the
resistivity is measured along the highly-conducting chains (or
layers), because there is no guarantee that the same chain (or
layer) is contacted by the four leads. Any interruption, stack
fault, kink, step, terraces, or other discontinuity causes
contributions of the interchain (or interlayer) resistivity. The
second problem common to organic but also inorganic crystals is
how to achieve low-resistance contacts which inject the current
homogeneously but do not cause cracks and non-linearities. For
that reason contactless microwave methods have been developed by
Schegolev {\it et al.} for investigating the electronic transport
in small and highly anisotropic crystals
\cite{Schegolev72,Helberg96}. It should be noted, that dc
measurements probe the voltage drop for a fixed electrical current
and thus are sensitive to any additional resistance added by
interchanging the conducting chains (or layers), for instance;
i.e.\ they measure the highest resistance for a given conductance
path. In contrast, microwave methods are sensitive to the current
induced by the electric field and thus probe the highest
conduction, weighted by the respective volume fraction, of course.
An interruption of a metallic sample by cracks, for instance, does
not influence the microwave properties since the high frequencies
easily bridge small gaps.

The microwave conductivity was obtained by the cavity-perturbation
method utilizing two different cylindrical copper cavities which
resonate in the TE$_{011}$ mode at 24 and 33.5~GHz, respectively.
They are fed by an Agilent E8257D analog signal generator via
suitable rectangular wave\-guides and operate in the transmission mode. The
coupling is about 10\%\ and done through two holes in the
sidewalls. The crystals are positioned in the maximum of the
electric field placed onto a quartz substrate (0.07~mm thick) in
such a way that the electric field is parallel to the $b$ axis.
The samples were cooled down slowly (0.2~K per minute)
from 300~K to 2~K by coupling to the liquid helium
bath with the help of low-pressure He exchange gas and by
utilizing a regulated heater. The stability is better than 10~mK
\cite{Dressel05}. By recording the center frequency $f$ and the
width $\Gamma$ (FWHM) of the resonance curve as a function of
temperature and comparing them to the corresponding parameters of
an empty cavity ($f_0$ and $\Gamma_0$), the complex electrodynamic
properties of the sample, like the surface impedance, the
conductivity and the dielectric constant, can be determined by
cavity-perturbation theory; further details on microwave
measurements and the data analysis are extensively discussed in
Ref.~\cite{Klein93,DresselGruner02}. Complementary ESR experiments have been performed down to $T=5$~K on \CHF\ using a Bruker X-band spectrometer.

\section{Results and Discussion}
\subsection{Organic Superconductor Close to Charge Ordering: \CHCF}
\label{sec:CHCF}
\begin{figure}
\centering\resizebox{0.4\textwidth}{!}{\includegraphics*{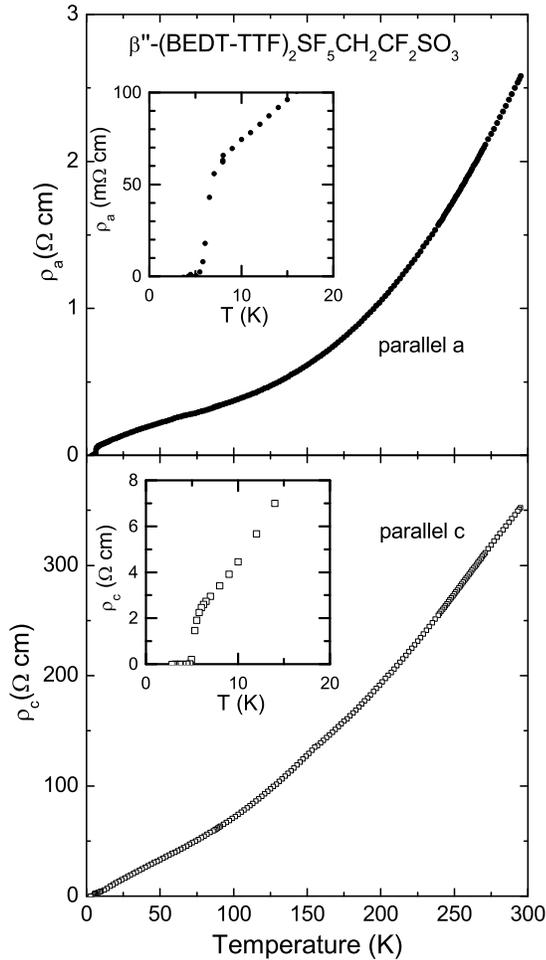}}
\caption{
\label{fig:CH2CF2_dc}
Temperature-dependent dc resistivity of \CHCF\ parallel and perpendicular to the highly conducting plane, i.e.\ parallel to the $a$ axis and parallel to the $c$ axis. The superconducting transition is observed at $T_c=5.9$~K.}
\end{figure}
In Fig.~\ref{fig:CH2CF2_dc} the in-plane and out-of-plane dc
resistivity of \CHCF\ is plotted as a function of temperature. The
absolute value of $\rho_a=2.6~{\rm \Omega cm}$ at room temperature is comparable to
the one reported in literature \cite{Dong99,remark2}. Previous
experiments \cite{Beckmann98,Su99,Hagel03,Hagel07} found a
metallic temperature dependence of the $c$-axis resistivity very
similar to our data, except for a higher absolute value. The room-temperature anisotropy within one crystal is ${\rho_c}/{\rho_a}
\approx 140$. Analyzing the asymmetry factor of the Dysonian line
shape obtained in X-band ESR experiments, Wang {\it et al.}
\cite{Wang99} estimated an in-plane anisotropy $\sigma_{\rm
max}/\sigma_{\rm min}$ of 1.35, with the maximum conductivity in
$b$ direction and $\sigma_{\rm min}$ basically parallel to the
stacks \cite{remark1}. This behavior is confirmed by optical
measurements \cite{Dong99,Kaiser08} and expresses the fact that
the orbital overlap between the stacks is stronger than along the stacks.
The resistivity ratio $R(300~{\rm K})/R(6~{\rm K})$ is
approximately 50 and 150 for the $a$ axis and 150 for interlayer transport.
At low temperatures, $\rho(T)$ follows a linear
behavior. Hagel {\em et al.}
\cite{Hagel07} report a $T^2$ behavior below 10~K which becomes linear above; the temperature region of the quadratic dependence could be extended upon applying external pressure.
The critical temperature of the superconducting
transition $T_c=5.9$~K  is
determined by the resistivity drop in both directions; its
comparably high value \cite{Geiser96,Ward00} evidences the
excellent quality of the single crystals.  As shown in the insets,
the width of the superconducting transition is $\Delta T_c \approx
1$~K.

It is interesting to note that for both directions the resistivity
$\rho(T)$ increases more or less linearly with temperature up to
approximately 125~K. It was theoretically predicted \cite{Merino06,Merino03} and for various $\alpha$-salts experimentally observed \cite{Drichko06} that the $T^2$ dependence of a regular metal changes to a linear temperature dependence of the scattering rate when charge-order fluctuations become important.
For $\rho_c(T)$ we may identify a second
change in slope around $T=200$~K. From optical experiments, we see first indications that the infrared-active vibrational features $\nu_{27}({\rm B_{1u}})$ split already
around 200~K \cite{Drichko08}; at $T=150$~K two Raman modes are clearly distinguished,
which evidences charge disproportionation \cite{Kaiser08}. Other indications of  charge fluctuations come from the electronic bands.

\begin{figure}
\centering\resizebox{0.4\textwidth}{!}{\includegraphics*{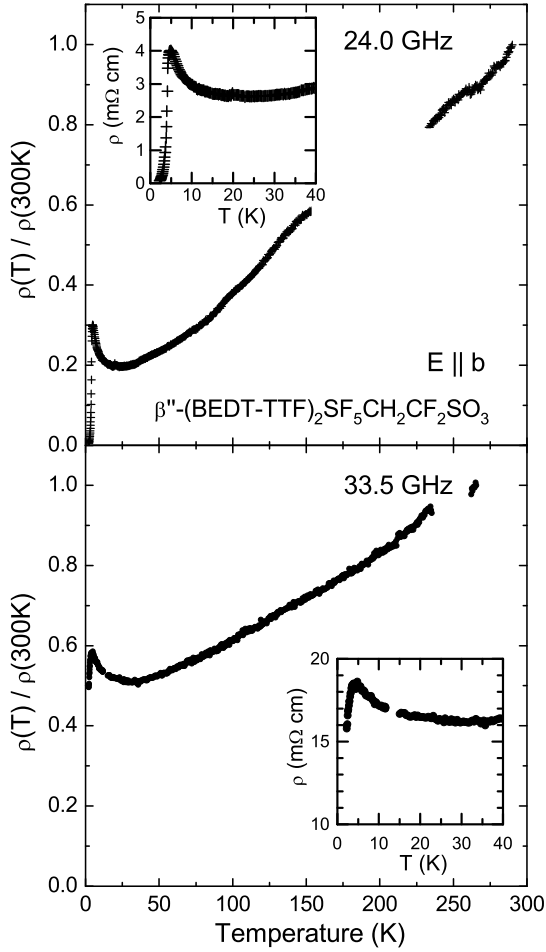}}
\caption{ \label{fig:CH2CF2_24+33} Temperature-dependent in-plane
resisitivity of \CHCF\ measured at 24 and 33.5~GHz normalized to
their room-temperature value. The inset shows the low-temperature
resistivity. For these specimen the resistivity increases below
approximately 30~K, before the superconducting transition is
observed.}
\end{figure}

The microwave properties obtained at $f=24.0$ and 33.5 GHz along
the $b$ axis are presented in Fig.~\ref{fig:CH2CF2_24+33}; the
data are normalized to the room-temperature value because the
uncertainty in absolute values is large (and can exceed a factor
of 10) due to errors in the depolarization factor caused by the
irregular shape of the samples. Nevertheless, from our data we
roughly estimate $\rho(300~{\rm K}) \approx 30~{\rm m \Omega\, cm}$ which is about one order of magnitude less than the dc values.
Starting at ambient temperature, the resistivity drops continuously
down to $T=30$~K as expected for a metal. This change is larger for
lower frequencies. A similar tendency is in general observed in
unconventional metals and indicates a strong frequency dependence
even in the microwave range.

In both microwave measurements, at 24 and 33.5~GHz, the
resistivity exhibits a minimum around $T=20$ to 30~K and turns up
by approximately 20 to 30\%\ when cooled down further; this
feature is robust. We relate the resistivity upturn to
charge-order fluctuations which may be even responsible for
superconductivity. From optical experiments (infrared and Raman
vibrational spectroscopy) it is known that \CHCF\ is subject to
charge ordering around 125-150~K \cite{Kaiser08}. From that we
would expect that a pseudogap opens which causes a reduction of
the density of states and hence a lower conductivity. However, the
coherent-particle zero-frequency transport is not affected; and
the low-frequency optical measurements also give clear evidence
for a Drude-like peak. Alternatively, the increased microwave
resistivity could be caused by enhanced scattering on charge
fluctuations. ESR experiments do not indicate any magnetic order
prior to the superconducting transition \cite{Wang99}.

The superconducting transition is observed at $T_c=5.0$~K for the
24~GHz experiment and around 4.5~K in the 33.5~GHz data. The lower
value compared to the dc measurement can be explained by the weak
thermal coupling of the sample in the microwave cavity.

Fermi-surface studies of \CHCF\ by Shubnikov-de Hass oscillations
see no indications of one-dimensional sheets;
they further give evidence that the two-dimensional part of the Fermi surface
contains only 5\%\ of the first Brillouin zone \cite{Beckmann98}.
This implies that the charge carriers available for dc transport
are significantly reduced, leading to a small spectral weight of
the Drude-like optical conductivity. From our high-frequency
measurements we can now conclude that already at 20 to 30~GHz the
roll-off in the frequency-dependent conductivity
is taken effect. The non-monotonous temperature behavior
infers charge fluctuations being responsible for this
behavior.

\subsection{Organic Superconductor: \d8CHCF}
\label{sec:d8CHCF} 
\begin{figure}
\centering\resizebox{0.4\textwidth}{!}{\includegraphics*{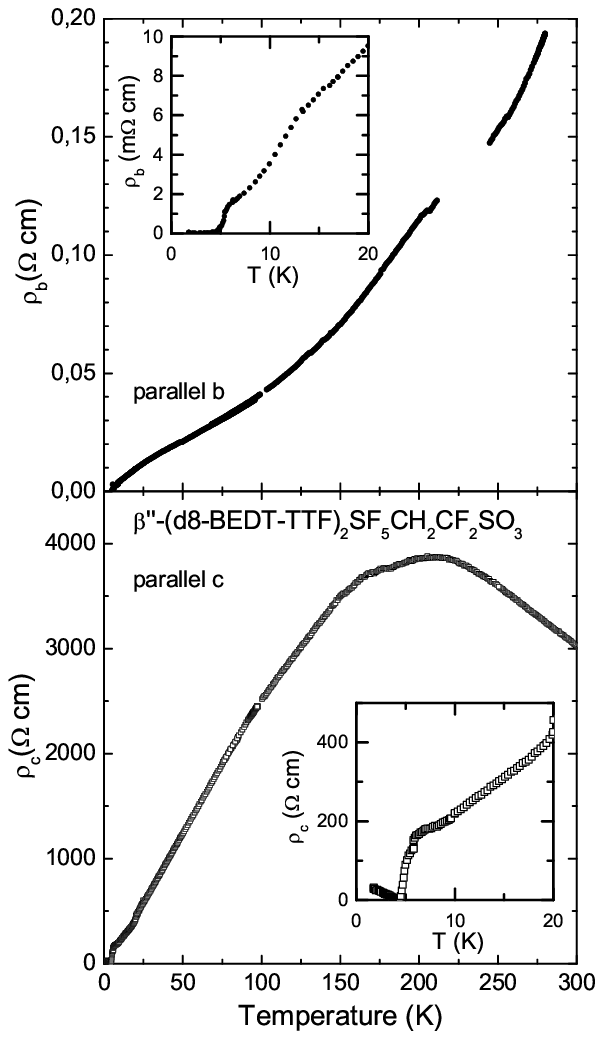}}
\caption{ \label{fig:d8CH2CF2_dc} Temperature-dependent
dc-resistivity of \d8CHCF\ parallel and perpendicular to the
highly conducting plane, i.e.\ parallel to the $b$ axis and
parallel to the $c$ axis. The superconducting transition occurs at
$T_c=5.6$~K.}
\end{figure}

The dc-transport properties of the deuterated
organic superconductor \d8CHCF\ are displayed in
Fig.~\ref{fig:d8CH2CF2_dc}; the current is either directed within
the  planes or perpendicular to them, as indicated. For the
highly-conducting $b$ direction a simple metallic response is
observed with a resistivity ratio $R(300~{\rm K})/R(6~{\rm K}) =
100$. The exponent $\alpha$ in the power law $\rho(T)\propto
T^{\alpha}$ is slightly below 2. Measurements of magnetic quantum
oscillations evidenced virtually no difference in the
low-temperature band-structure parameters compared to the
hydrogenated sister compound. At $T_c=5.6$~K the onset of the
superconducting transition is observed which is about 1~K broad; a
somewhat lower critical temperature was reported previously based
on ac-susceptibility measurements \cite{Schlueter01}. In the $c$
direction the resistivity is very large indicating the pronounced
two-dimensionality of the material. Cooling down from ambient
temperature the resistivity increases first, goes through a
maximum around $T=200$~K and then continuous to drop down to
lowest temperatures in a linear fashion, before it exhibits
superconductivity \cite{remark3}.

\begin{figure}
\centering\resizebox{0.4\textwidth}{!}{\includegraphics*{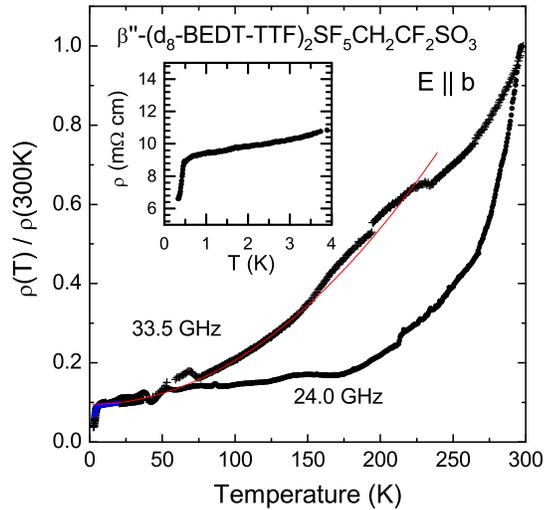}}
\caption{ \label{fig:d8CH2CF2_24+33} Temperature-dependent
resistivity of \d8CHCF\ measured at 24 and 33.5~GHz. The critical
temperature for the superconducting phase is $T_c=5.6$~K. The
solid line indicates a fit to a quadratic behavior. The inset
exhibits a low-temperature blow-up of the 24-GHz data.}
\end{figure}

Fig.~\ref{fig:d8CH2CF2_24+33} shows the microwave resistivity of
\d8CHCF\ measured along the highly conducting $b$ direction at
$24$ and 33.5 GHz. At room temperature the absolute values are
approximately $0.1~\rm \Omega$cm; i.e.\ they are comparable to the
dc resistivity. While the resistivity ratio is similar in both
experiments [$\rho(300~{\rm K})/\rho(6~{\rm K})$ $\approx 10$], at
$f=24$~GHz we observe a dramatic change at temperatures above
200~K while the higher frequency data drop gradually in a more or
less quadratic temperature behavior. This confirms the general
behavior already mentioned for \CHCF, that chang\-es in
resistivity with temperature become smeared out as the measurement
frequency increases. The transition to the superconducting phase
is detected at 5.6~K. The microwave resistivity flattens out below
50~K and exhibits only a very weak temperature dependence. But
most important, there are no indications of a resistivity upturn
as observed in the hydrogenated analogue (cf.\
Fig.~\ref{fig:CH2CF2_24+33}). With other words, since the
deuterated crystals remain metallic even in their microwave
properties, they seem to be less susceptible to charge
fluctuations.

A careful investigation \cite{Schlueter01} of the isotope shift in
a large number of BEDT-TTF samples revealed that for \CHCF\ the
deuteration causes an increase of the superconducting transition
temperature by about 0.3~K. From pressure-dependent measurements
\cite{Hagel03,Sadewasser97} we know that the superconducting
transition temperature decrease with pressure by a rate of -1.3
K/bar. This is due to the weaker influence of the effective
intersite Coulomb repulsion $V/t$ as $t$ increases with pressure;
the systems becomes more metallic \cite{remark5}. The increase of
$T_c$ would therefore infer that \d8CHCF\ is closer to the
charge-order transition. However, due to the known
sample-to-sample variation, we cannot make any final statement in
this regard and have to leave this discrepancy unresolved.

\subsection{Organic Metal: \CHF}
\label{sec:CHF}
\begin{figure}
\centering\resizebox{0.45\textwidth}{!}{\includegraphics*{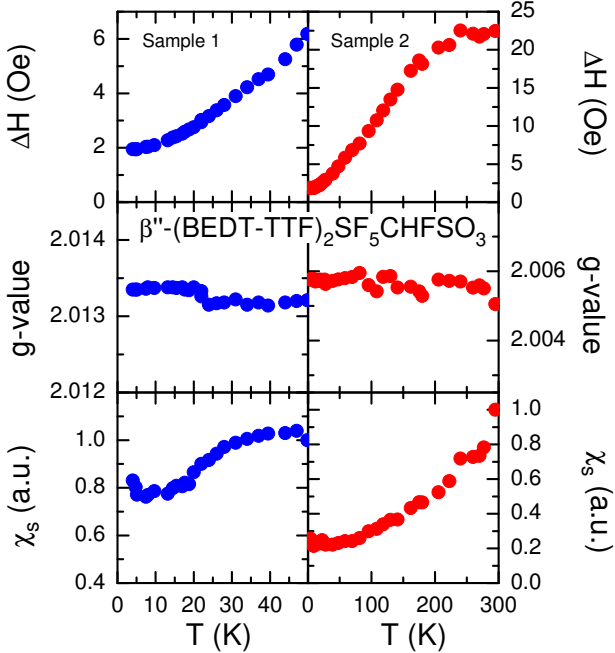}}
\caption{ \label{fig:CHF_esr} ESR data obtained from
temperature-dependent X-band measurements on two different \CHF\
samples. The panels show the linewidth $\Delta H$, the $g$-value
and the spin susceptibility $\chi_s(T)$ for Sample 1 (left column)
and Sample 2 (right column).}
\end{figure}

Replacing the (SF$_5$CH$_2$CF$_2$SO$_3)^-$ anions in the organic crystals by (SF$_5$CHFSO$_3)^-$ leads to a non-superconducting compound.
\begin{figure}
\centering\resizebox{0.4\textwidth}{!}{\includegraphics*{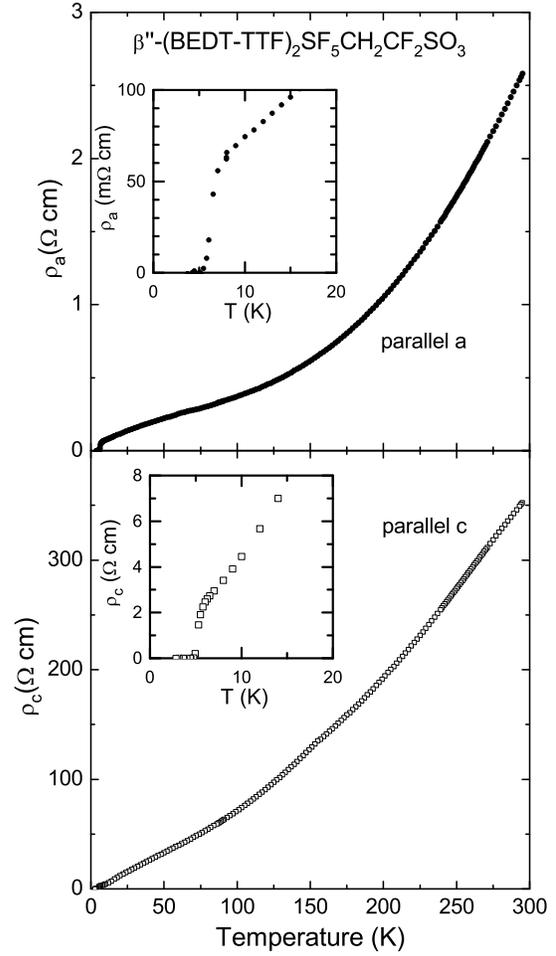}}
\caption{
\label{fig:CHF_dc}
Temperature-dependent dc-resistivity of \CHF\ parallel and perpendicular to the highly conducting plane; i.e.\ the along $b$ and $c$ axes, respectively. For most samples, a metallic behavior is observed down to low temperatures. Below approximately 50~K, indications of a metal-to-insulator transition are seen along the $c$ direction.}
\end{figure}
From room temperature down to approximately 50~K, the electrical
resistivity of \CHF\ exhibits a metallic temperature dependence in
both directions, parallel and perpendicular to the highly
conducting ($ab$) plane. The temperature dependences are presented
in Fig~\ref{fig:CHF_dc}. Along the $b$ axis, the absolute value
$\rho_b(300~{\rm K}) = 2.3~{\rm \Omega cm}$ is comparable to the
resistivity measured for \CHCF; the ratio $\rho_b(300~{\rm
K})/\rho_b(30~{\rm K}) = 15.3$. Below 50~K the resistivity seems
to saturate. A rise of the low-temperature resistivity was
reported by Ward {\it et al.} \cite{Ward00,Jones00} without
providing an explanation. They also mention that the spin
susceptibility slightly increases below 12~K which indicates that
electron localization takes place at very low temperatures. We
cannot confirm this finding from X-band ESR measurements on our
samples, as demonstrated in Fig.~\ref{fig:CHF_esr}. In the
perpendicular direction, $\rho_c(T)$ also shows a decrease in
resistivity when cooled down, as plotted in Fig~\ref{fig:CHF_dc}b.
The room-temperature anisotropy $\rho_c/\rho_b$ of \CHF\ is
approximately 200, i.e.\ somewhat higher than for \CHCF. For
$T<50$~K an insulating behavior is observed with a rapid increase
in resistivity well above the room-temperature value; no simple
thermally activated behavior can be determined; nevertheless the
initial slope corresponds to an activation energy $E_a=25$~meV,
when fitted by $\rho_c(T)\propto \exp\{E_a/k_B T\}$.

\begin{figure}
\centering\resizebox{0.4\textwidth}{!}{\includegraphics*{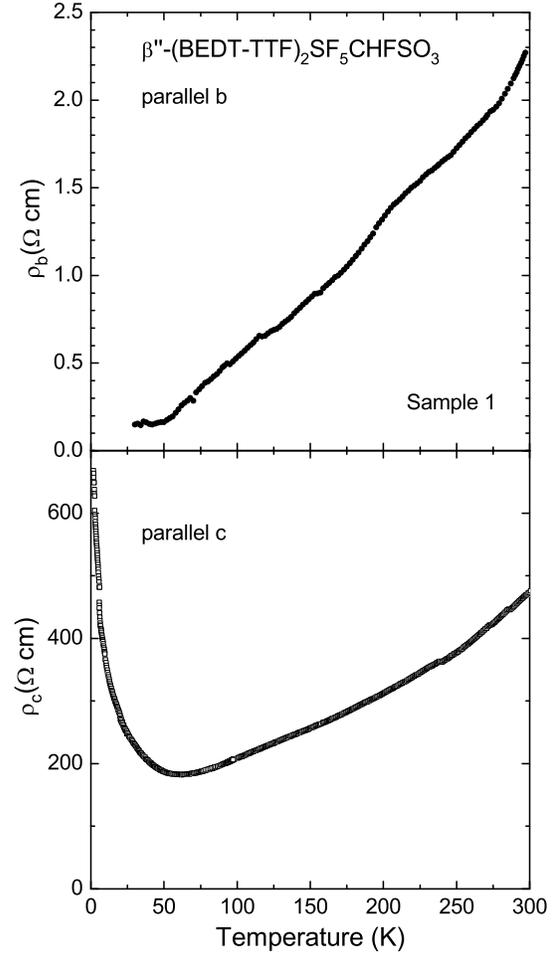}}
\caption{ \label{fig:CHF_dc2} Some of the \CHF\ samples exhibit an
insulating behavior in $\rho(T)$. The resistivity increases to a
maximum around 70~K, below which it decreases with a kink at
16~K.}
\end{figure}
It should be noted that for some of the samples (two out of eight)
a different temperature behavior of the in-plane resistivity was
measured which resembles previous reports \cite{Ward00}. Starting
from approximately the same am\-bi\-ent-temperature value,
$\rho_b(T)$ decreases slightly when the sample is cooled down to
250~K. Then the electrical resistivity becomes semiconducting and
gradually increases by about a factor of 4 all the way down to
70~K where a broad maximum if found (Fig~\ref{fig:CHF_dc2}). While
Ward {\em et al.} \cite{Ward00} could identify an activated
behavior between $T=100$ and 300~K with $E_a=56$~meV, the increase
we observe does not correspond to a single activation energy; only
between 150 and 200~K the behavior may be described by
$\rho_b(T)\propto \exp\{E_a/k_B T\}$ with $E_a=26$~meV. The drop
in $\rho_b(T)$ observed at lower temperatures exhibits a kink
around 16~K which might be an indication of a phase transition. At
this point we do not have a definite explanation, but the absence
of any change in the ESR data around this temperature
(Fig.~\ref{fig:CHF_esr}) is an argument against magnetic order
\cite{remark4}.

\begin{figure}
\centering\resizebox{0.4\textwidth}{!}{\includegraphics*{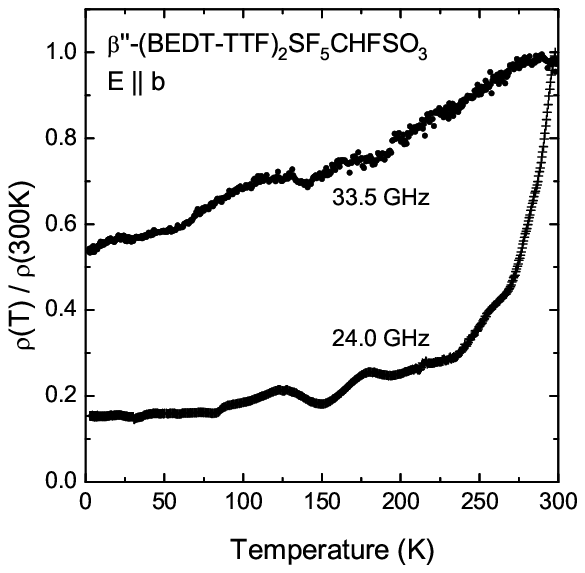}}
\caption{
\label{fig:CHF_24+33}
Temperature-dependent resisitivity of \CHF\ measured at 24 and 33.5~GHz. The data are normalized to the respective room-temperature value.}
\end{figure}

Microwave measurements of \CHF\ in the highly-conducting plane
reveal a metallic behavior down to lowest temperatures
(Fig.~\ref{fig:CHF_24+33}). No indication of charge order could be
detected. Again, for $f=33.5$~GHz the temperature dependence of
the resistivity changes more gradually (almost linearly), while
for 24~GHz we observe a fast drop below room temperature with only
little variation for $T<150$~K. At low temperature we did not find
an upturn in resistivity as present in the dc data (cf.\
Fig.~\ref{fig:CHF_dc}). This basically implies that the
semiconducting behavior  shows up only for $\omega\rightarrow 0$
while in ac-transport and optical experiments the charge
localization can be overcome.

It should be pointed out that for the superconducting crystals \CHCF\ the tendency was opposite: there charge fluctuations led to a rise in high-frequency resistivity upon cooling below $T=20$~K prior to the superconducting transition (see Sec.~\ref{sec:CHCF} and in particular Fig.~\ref{fig:CH2CF2_24+33}) while the dc resistivity remains purely metallic. In the present case of \CHF, we feel that the strongly insulating behavior observed in the perpendicular direction (Fig.~\ref{fig:CHF_dc}) also affects $\rho_b$ because due to unavoidable crystal imperfections (e.g.\ cracks) the current path might have to change planes. This is not the case for  microwave measurements which are not sensitive to these imperfections and probe the electrical conduction more directly as discussed in Sec.~\ref{sec:experimental}.

\section{Conclusions}
A comprehensive investigation of transport properties of different
$\beta^{\prime\prime}$-phase organic conductors and
superconductors has been presented, utilizing dc and microwave
methods at different frequencies (24 and 33.5~GHz) from room
temperature down to 2~K. At ambient conditions the in-plane dc
conductivity is of the order $1~({\rm \Omega cm})^{-1}$; the
an\-iso\-tropy $\sigma_{\parallel}/\sigma_{\perp}$ exceeds $10^2$.
Hence, the systems are highly anisotropic two-dimensional metals.
All compounds exhibit a more or less metallic behavior upon
cooling with \CHCF\ and the deuterated analog \d8CHCF\ becoming superconducting around
5~K. For \CHCF\ clear signatures of charge fluctuations are
present for $T<30$~K. While they do not influence the dc
properties,  they lead to an increase of the high-frequency
resistivity. These effects are not seen in the deuterated
analog, maybe due to the better metallic behavior.

Considering the frequency dependence of the conductivity as
derived from our experiments, the general behavior certainly
deviates from that of simple metals. We observe a more gradual
temperature dependence for higher frequencies, indicating a
significant frequency dependence of the electrodynamic response
even in the microwave range. Since this happens at elevated
temperatures ($T>100$~K), we can rule out effects due to
conventional phonon scattering. Instead we are inclined to trace
it back to the peculiar bandstructure and correlation effects. The
difference in frequency dependence is strongest in \CHF\ and
weakest in \CHCF. Our investigations provide evidence that the
vicinity to charge order and the presence of charge
fluctuations strongly alter the dynamical response of the
electronic system.

\section*{Acknowlegements}
The dc measurements were done with the support of E.\ Rose. We would
like to thank G. Untereiner for the careful sample preparation.
The work at Stuttgart was supported by the Deutsche
Forschungsgemeinschaft (DFG). The crystal growth at Argonne
National Laboratory was performed under the auspices of the Office
of Basic Energy Sciences, Division of Material Sciences of the
U.S. Department of Energy, Contract No.\ DE-AC02-06CH11357.


\begin{thebibliography}{99}
\bibitem{remark1}The crystal symmetry is triclinic, with the room-temperature unit-cell parameters $a = 9.260$~\AA, $b=11.625$~\AA, $c=17.572$~\AA, $\alpha = 94.69^{\circ}$, $\beta=91.70^{\circ}$, $\gamma=103.10^{\circ}$, $Z=2$, the space group is $P{\bar 1}$.
\bibitem{remark3}Since the sample exhibited cracks upon cooling, some residual resistance remains in the
superconducting phase.
\bibitem{remark5}The insulating state found at even higher pressure \cite{Hagel03} is probably caused by some structural change and not due to electronic correlations.
\end{thebibliography}
\end{document}